\documentclass[12pt]{article}
\usepackage{graphicx}
\usepackage{amsmath}
\usepackage{amsfonts}
\usepackage{amssymb}
\def\const{{\rm const}}
\begin{document}
\title{Snyder-de Sitter model from two-time physics}
\author{M.C. Carrisi$^{1,*}$ and S. Mignemi$^{1,2,\dag}$}
\date{}
\maketitle
\small {\em \center $^{1}$Dipartimento di Matematica ed Informatica,
Universit\`{a} di Cagliari, viale Merello 92,\,\ 09123 Cagliari, Italy\\
$^{2}$INFN, Sezione di Cagliari, Cagliari, Italy\\ 
$^{*}$e-mail: cristina.carrisi@tiscali.it\\
$^{\dag}$e-mail: smignemi@unica.it\\}
\begin{abstract}
\noindent We show that the symplectic structure of the Snyder model on a de Sitter 
background
can be derived from two-time physics in seven dimensions and propose a Hamiltonian 
for a free particle consistent with the symmetries of the model.
\end{abstract}

\section{Introduction}
Some time ago, Romero and Zamora \cite{RZ} derived the phase space of the Snyder 
model \cite{sny} from the two-time (2T) physics model introduced in ref.\ 
\cite{bar}\footnote{In a different context, a similar model was 
also discussed in \cite{mar}.}.

Two-time physics is a $D$-dimensional model defined on phase space, 
with two timelike dimensions,
having the $SO(D-2,2)$ group as a global symmetry and the two-dimensional 
symplectic group $Sp(2)$ as the local one, whose Hamiltonian is given by a 
combination of homogeneous quadratic constraints in the phase space variables.
The introduction of two coordinates with timelike signature is necessary in 
order to satisfy the constraints in a nontrivial way.
By fixing the gauge freedom, one can recover several well-known four-dimensional 
models, like the massive relativistic particle in flat spacetime, or the massless 
particle in de Sitter spacetime. 
This formalism makes manifest some hidden symmetries of ordinary physics and can
be extended to include background gauge fields of any spin and noncommutative 
spacetime \cite{bar2}.

As for the Snyder model, it was introduced long time ago, in order to show the 
possibility of defining a noncommutative spacetime invariant under the action of the 
Lorentz group. 
Later, it was  interpreted as an instance of doubly special relativity \cite{KN}. 
This is a theory admitting two fundamental scales \cite{ame}, usually identified with 
the speed of light and the Planck mass. The presence of two fundamental scales 
enforces a deformation of the spacetime symmetries, in particular of translation 
invariance \cite{Mi1}.

More recently, the Snyder model has been generalized to the case of a de Sitter background 
\cite{KS,Mi2}. The resulting Snyder-de Sitter (SdS) model, called triply special relativity
by the authors of ref.\ \cite{KS} because of the presence of the cosmological constant as a 
third independent fundamental scale, displays a duality between position and momentum 
coordinates. 

Although the generalization of the Snyder model to a de Sitter background is
not unique \cite{Mi2}, a notably elegant extension has been proposed in ref.\ \cite{KS}.
The fundamental Poisson brackets postulated in \cite{KS} read
\begin{eqnarray}\label{sds}
  \{X_\mu,X_\nu\}&=&-\frac{1}{\kappa^2}(X_\mu P_\nu-X_\nu P_\mu), \nonumber\\
  \{P_\mu,P_\nu\}&=&-\frac{1}{\alpha^2}(X_\mu P_\nu-X_\nu P_\mu), \nonumber\\
  \{X_\mu,P_\nu\}&=&\eta_{\mu\nu}-\frac{1}{\alpha^2}X_\mu X_\nu-\frac{1}{\kappa^2}P_\mu P_\nu
  -\frac{2}{\alpha\kappa}P_\mu X_\nu,
\end{eqnarray}
with $\kappa$ the Planck energy and $\alpha$ the de Sitter radius. As discussed in \cite{CO}, 
this specific realization of the SdS model turns out to be a nonlinear realization of the
Yang model \cite{Ya}, proposed by Yang soon after Snyder's paper.

Previously, a derivation of the SdS model from a six-dimensional system with Lorentzian 
signature and nonhomogeneous constraints has been given in \cite{Mi3}. In this paper we 
provide an alternative derivation from a different higher-dimensional system, showing that 
the SdS model can be obtained starting from 7-dimensional two-time physics.

\section{The 2T model}
The 2T model \cite{bar} is defined on a flat $D$-dimensional manifold with two timelike 
coordinates
and signature $(+,-,\dots,-,+)$. Its action can be written as\footnote{Capital latin indices 
run from 0 to $D-1$. We denote $V^2\equiv V_A V^A$, $V\cdot W\equiv V_A W^A$. In the following, 
greek indices run from 0 to 3 and $V_\rho^2\equiv V^\rho V_\rho$.} \cite{bar}
\begin{equation}\label{1}
S=\int\left[\dot{X}\cdot P-\left(\lambda_1 \frac{1}{2}P^2+\lambda_2
X\cdot P+\lambda_3 \frac{1}{2}X^2 \right)\right]d\tau,
\end{equation}
and is invariant under the global $SO(D-2,2)$ symmetry with generators
 $J_{AB}\equiv X_AP_B-X_BP_A$.
The Hamiltonian is therefore
\begin{equation}\label{2}
H=\lambda_1 \frac{1}{2}P^2+\lambda_2 X\cdot P+\lambda_3 \frac{1}{2}X^2,
\end{equation}
with $\lambda_1$, $\lambda_2$ and $\lambda_3$ Lagrange multipliers that enforce the constraints
\begin{eqnarray}\label{3}
  \phi_1&=&\frac{1}{2}P^2 \approx 0,\nonumber\\
  \phi_2&=&X\cdot P \approx 0,\\
  \phi_3&=&\frac{1}{2}X^2 \approx 0,\nonumber
\end{eqnarray}
where the weak equivalence is used in the sense of Dirac \cite{dir}.
The Hamilton equations following from (\ref{1}) are
\begin{eqnarray}\label{4}
\dot X_A&=&\{X_A,H\}=\lambda_1P_A+\lambda_2X_A, \nonumber\\
\dot P_A&=&\{P_A,H\}=-\lambda_2P_A-\lambda_3X_A.
\end{eqnarray}
No secondary constraints are present, since
\begin{eqnarray*}
 \dot{\phi_1}&=&\{\phi_1,H\}=-2\lambda_2\phi_1-\lambda_3\phi_2 \approx 0, \\
 \dot{\phi_2}&=&\{\phi_2,H\}=-2\lambda_1\phi_1-2\lambda_3\phi_3 \approx 0, \\
 \dot{\phi_3}&=&\{\phi_3,H\}=\lambda_1\phi_2+2\lambda_2\phi_3 \approx 0.
\end{eqnarray*}

The Poisson brackets between the constraints generate the $sp(2)$ algebra,
\begin{equation}\label{5}
\{\phi_1,\phi_2\}=-2\phi_1, \qquad  \{\phi_1,\phi_3\}=-\phi_2, \qquad
\{\phi_2,\phi_3\}=-2\phi_3,
\end{equation}
and all the constraints are therefore first class. Due to the presence of three 
first-class constraints, the original $2D$ coordinates of the phase space  
reduce after gauge fixing to $2(D-3)$ independent ones.

\section{Snyder space}
In constrained Hamiltonian systems, the presence of arbitrary functions $\lambda_i$ in 
the Hamiltonian indicates that the correspondence between the physical states and the 
canonical variables is not one to one, but to a given state can correspond different 
sets of values of the canonical variables. 
This redundance is called gauge invariance and the transformations that connect different 
sets of equivalent variables are generated by the first class constraints \cite{dir}. 
The problem can be solved by imposing new constraints, called gauge conditions, 
that reduce the first class constraints to second class, and decrease the
number of independent variables, restoring a one-to-one correspondence between the 
physical states and the independent phase space coordinates.

In our case, the choice of specific gauge conditions leads to different lower-dimensional 
models. In particular, the authors of \cite{RZ} showed that for $D=6$, the choice
\begin{equation}\label{6a}
P_4=L=\const, \qquad \qquad X_4=0,
\end{equation}
reduces the dynamics to that of a four-dimensional particle with independent variables 
$X_\mu$ and $P_\mu$, while $X_5$ and $P_5$ become functions of the other variables,
\begin{equation}
P_5=\sqrt{L^2-P_\rho^2},\qquad X_5=\frac{-P_\sigma X^\sigma}{\sqrt{L^2-P_\rho^2}}.
\end{equation}

After imposing the gauge constraints (\ref{6a}), two of the constraints (\ref{3}) become
second class. For consistency,
the gauge choice  must be preserved under the evolution of the system. This 
can be achieved if one imposes $\lambda_1=\lambda_2=0$, which is in accordance with the 
fact that after the elimination of $X_4$, $X_5$, $P_4$ and $P_5$, one is left with 8 
degrees of freedom and a single first-class (Hamiltonian) constraint, given by
\begin{equation}\label{7a}
H=\frac 1 2\left[\frac{(P_\sigma X^\sigma)^2}{L^2-P_\rho^2}+X_\rho^2\right]=0.
\end{equation}
Some drawbacks are however present in the definition of the Hamiltonian constraint.
First of all, it does not look very attractive, although it can be written in the
equivalent form
\begin{equation}
\left(\eta^{\mu\nu}-\frac{X^\mu X^\nu}{X_\rho^2}\right) P_\mu P_\nu=L^2,
\end{equation}
resembling that of a massive particle. Moreover,
the equations obeyed by the independent variables $X_\mu$ and $P_\mu$ are 
\begin{equation}
\dot X_\mu=0,\qquad\dot P_\mu=-\lambda_3X_\mu,
\end{equation}
and do not seem to have a sensible physical interpretation.

One can easily derive the Dirac brackets satisfied by the phase space variables after the
elimination of the second class constraints,
\begin{eqnarray}\label{11a}
  \{X_\mu,X_\nu\}^*&=&-\frac{1}{L^2}(X_\mu P_\nu-X_\nu P_\mu), \nonumber\\
  \{P_\mu,P_\nu\}^*&=& 0, \nonumber\\
  \{X_\mu,P_\nu\}^*&=&\eta_{\mu\nu}-\frac{1}{L^2}P_\mu P_\nu.
\end{eqnarray}
These are precisely the commutation relations of Snyder space \cite{sny}. Had one chosen the
alternative gauge $P_5=L$, $X_5=0$, one would have recovered instead the 
so-called anti-Snyder model, which obeys the same Poisson brackets, with $L^2\to-L^2$.

By duality, a similar calculation with the gauge choice 
\begin{equation}\label{6b}
X_4=M=\const, \qquad \qquad P_4=0,
\end{equation}
gives rise to a model with Dirac brackets for the phase space variables identical to
the fundamental Poisson brackets of a free massless particle in de Sitter spacetime \cite{mar},
\begin{eqnarray}\label{11b}
  \{X_\mu,X_\nu\}^*&=&0, \nonumber\\
  \{P_\mu,P_\nu\}^*&=&-\frac{1}{M^2}(X_\mu P_\nu-X_\nu P_\mu), \nonumber\\
  \{X_\mu,P_\nu\}^*&=&\eta_{\mu\nu}-\frac{1}{M^2}X_\mu X_\nu,
\end{eqnarray}
In this case, consistency requires $\lambda_2=\lambda_3=0$, and 
the Hamiltonian constraint reads
\begin{equation}\label{7b}
H=\frac 1 2\left(\eta^{\mu\nu}+\frac{X^\mu X^\nu}{M^2-X_\rho^2}\right) P_\mu P_\nu=0,
\end{equation} 
which is proportional to that of a massless particle in de Sitter spacetime with cosmological 
constant $M$, in stereographic coordinates. The equations obeyed by the independent variables
are
\begin{equation}
\dot X_\mu=\lambda_1P_\mu,\qquad\dot P_\mu=0.
\end{equation}
Also in this case, the Hamiltonian and the equations of motion are not the standard ones for
de Sitter space.

Finally we notice that, in analogy with the previous case, the gauge choice 
$X_5=M$, $P_5=0$, would lead to anti-de Sitter spacetime.

\section{The SdS model}
From the previous results, one may guess that the SdS model can be obtained from the two-time 
model in a similar way. However, it turns out that the Poisson brackets of
the SdS model can only be obtained starting from $D=7$. The fixing of all the gauge degrees
of freedom will then lead to a 8-dimensional phase space, and a further Hamiltonian constraint
must be imposed if one wants to describe the dynamics of the 4-dimensional SdS particle.

It is natural to consider the following gauge conditions:
\begin{equation}\label{6c}
P_4=L=\const, \qquad \qquad X_4=M=\const.
\end{equation}
Moreover, as a third gauge condition, we choose
\begin{equation}\label{7c}
M P_5+L X_5=0.
\end{equation}
In this way, we have fixed all the gauge freedom, and the Lagrange multipliers $\lambda_i$
must vanish for consistency with (\ref{4}). 
From the constraints and the gauge conditions it follows that
\begin{eqnarray}\label{14}
X_5&=&\pm M\sqrt{\frac{X_\mu^2P_\nu^2-(X^\mu P_\mu)^2-(MP_\mu-LX_\nu)^2}{4L^2M^2-(M P_\rho+L X_\rho)^2}},
\nonumber\\
P_5&=&\mp L\sqrt{\frac{X_\mu^2P_\nu^2-(X^\mu P_\mu)^2-(MP_\mu-LX_\nu)^2}{4L^2M^2-(M P_\rho+L X_\rho)^2}},
\nonumber\\
X_6&=&\pm\frac{2M^2L-LX_\mu^2-MX^\mu P_\mu}{\sqrt{4L^2M^2-(M P_\rho+L X_\rho)^2}},\nonumber\\
P_6&=&\pm\frac{2ML^2-MP_\mu^2-LX^\mu P_\mu}{\sqrt{4L^2M^2-(M P_\rho+L X_\rho)^2}}.
\end{eqnarray}
In this way we have fully reduced the system to a 8-dimensional one, spanned by the coordinates
$X_\mu$ and $P_\mu$.

Let's now consider the constraints $(\ref{3})$ together with the gauge constraints
\begin{eqnarray}\label{8}
\chi_1&=&P_4-L ~~\approx 0, \nonumber \\
\chi_2&=&X_4-M \approx 0, \nonumber \\
\chi_3&=&M P_5+L X_5\approx 0,
\end{eqnarray}
and calculate their Poisson brackets. We obtain
\begin{eqnarray}\label{XP}
 && \{\chi_1,\chi_2\}=1, \qquad \{\chi_1,\chi_3\}=0, \qquad \{\chi_2,\chi_3\}=0\nonumber \\
 && \{\phi_1,\chi_1\}=0, \qquad  \{\phi_1,\chi_2\}=-P_4, \qquad \{\phi_1,\chi_3\}=-LP_5,\nonumber \\
 && \{\phi_2,\chi_1\}=P_4, \qquad  \{\phi_2,\chi_2\}=-X_4, \qquad \{\phi_2,\chi_3\}=MP_5-LX_5,\nonumber \\
 && \{\phi_3,\chi_1\}=X_4, \qquad  \{\phi_3,\chi_2\}=0, \qquad \{\phi_3,\chi_3\}=MX_5.
\end{eqnarray}
It follows that all constraints are now second class.
Their Poisson brackets are encoded in the following matrix:
\begin{equation}\label{10}
C_{\alpha \beta}\equiv\{\chi_\alpha,\chi_\beta\}=\begin{pmatrix}
  0 & 0 & 0 & 0 &-L &-LP_5 \\
  0 & 0 & 0 & L &-M &2MP_5 \\
  0 & 0 & 0 & M & 0 & -M^2P_5/L \\
  0 &-L &-M & 0 & 1 & 0 \\
  L & M & 0 &-1 & 0 & 0 \\
  LP_5 &-2MP_5 & M^2P_5/L & 0 & 0 & 0 \nonumber
\end{pmatrix},
\end{equation}
whose inverse is
\begin{equation}\label{10b}
C^{\alpha \beta}=\frac{1}{4L^2M^2}\begin{pmatrix}
  0 & M^2 & 2LM & M^3 & 3LM^2 & LM^2/P_5 \\
 -M^2 & 0 & L^2 &-LM^2 & ML^2 &-L^2M/P_5 \\
 -2LM &-L^2 & 0 &-3ML^2 &-L^3 & L^3/P_5 \\
 -M^3 & LM^2 & 3ML^2 & 0 & 0 & 0 \\
 -3LM^2 &-ML^2 & L^3 & 0 & 0 & 0 \\
 -LM^2/P_5 & L^2M/P_5 &-L^3/P_5 & 0 & 0 & 0 \nonumber
\end{pmatrix}.
\end{equation}
The Dirac brackets for the phase space coordinates $X_\mu$, $P_\mu$,
defined as $\{A,B\}^*=\{A,B\}-\{A,\chi_\alpha\}C^{\alpha\beta}\{\chi_\beta,B\}$,
are then given by
\begin{eqnarray}\label{11}
  \{X_\mu,X_\nu\}^*&=&-\frac{1}{4L^2}(X_\mu P_\nu-X_\nu P_\mu), \nonumber\\
  \{P_\mu,P_\nu\}^*&=&-\frac{1}{4M^2}(X_\mu P_\nu-X_\nu P_\mu), \nonumber\\
  \{X_\mu,P_\nu\}^*&=&\eta_{\mu\nu}-\frac{1}{4M^2}X_\mu X_\nu-\frac{1}{4L^2}P_\mu P_\nu-
  \frac{1}{2LM}P_\mu X_\nu.
\end{eqnarray}
These are identical to the Poisson brackets (\ref{sds}) for $M=\alpha/2$, 
$L=\kappa/2$.

In order to define the dynamics, one must now add a further constraint, which corresponds
to the Hamiltonian constraint of the ordinary relativistic particle. This was not necessary in
the case of the flat Snyder model because, due to the lower dimensionality, one constraint was 
left after the reduction from six to four dimensions.

The most natural choice is the quadratic Casimir invariant of the residual $SO(1,4)$ symmetry
generated by $J_{\mu\nu}$ and $J_{\mu6}$. This is given by
\begin{equation}
H_4=J_{\mu\nu}^2+2J_{\mu6}^2=N^2,
\end{equation}
where $N$ is a constant, proportional to the mass of the particle.
Using (\ref{14}), one obtains more explicitly, modulo a constant factor,
\begin{equation}\label{Ham}
H_4=LM\, \frac{M^2P_\mu^2+L^2X_\mu^2-X_\mu^2P_\nu^2+(X^\mu P_\mu)^2-2LMX^\mu P_\mu}
{4L^2M^2-M^2P_\mu^2-L^2X_\mu^2-2LMX^\mu P_\mu}.
\end{equation}
In spite of the ugly expression of the Hamiltonian, the Hamilton equation derived from 
(\ref{Ham}) with the help of the brackets (22), take a very simple form
\begin{eqnarray}
\dot X_\mu&=&-{1\over2L}\left[(1-N^2)LX_\mu-(1+N^2)MP_\mu\right],\cr
\smallskip\cr
\dot P_\mu&=&-{1\over2M}\left[(1+N^2)LX_\mu+(1-N^2)MP_\mu\right].
\end{eqnarray}
In second order form, using
\begin{equation}
P_\mu = {1\over1+N^2}\,{L\over M}\,[2\dot X_\mu-(1-N^2)X_\mu],
\end{equation}
they reduce to
\begin{equation}
\ddot X_\mu=-N^2\,X_\mu.
\end{equation}
Hence, each position coordinate satisfies the equation of a harmonic oscillator 
(or a free particle in the massless case).

A different possibility is to choose the Hamiltonian constraint like in de Sitter 
space, as proposed in \cite{KS},
\begin{equation}\label{Ham2}
(4M^2-X_\mu^2)P_\nu^2+(X^\mu P_\mu)^2=N^2.
\end{equation}
In this case the Hamilton equations are
\begin{equation}
\dot X_\mu={2\over L^2}\big[4L^2M^2-M^2P_\rho^2-L^2X_\rho^2-2LMX^\rho P_\rho\big]P_\mu,
\qquad \dot P_\mu=0.
\end{equation}
and the momentum $P_\mu$ is conserved, while the coordinates $X_\mu$ satisfy coupled
first order equations.
However, the Hamiltonian breaks the symmetry for the interchange of $P_\mu$ with $X_\mu$
and therefore looks less natural than (\ref{Ham}).

\section{Conclusions}
We have shown that the phase space of the SdS model can be realized starting from the 
7-dimensional 2T model. Contrary to the derivation of the flat space Snyder model, the
Hamiltonian constraint is not included in the original constraints, but must be added
by hand. This fact leaves a greater freedom in the choice of the dynamics, avoiding the
problems found in ref.\ \cite{RZ}, and allowing the introduction of massive particles.
In particular, it is possible to choose a Hamiltonian that preserves the duality
invariance for the interchange of $X_\mu$ and $P_\mu$.

A different derivation of the SdS model from higher dimensions was proposed in \cite{Mi3},
starting from a 6-dimensional model with Lorentz signature and inhomogeneous constraints. 
The possibility of using a lower dimensionality in this case, is due to the inhomogeneity 
of the constraints, that reduces the local symmetry group to $U(1)$ instead of $Sp(2)$. 
In that derivation, however, the values of $\alpha$ and $\kappa$ are no longer free, but 
have to be fixed from the beginning.

\end{document}